\documentclass[letterpaper]{article}

\title{An analysis of the Act 43 Wisconsin Assembly district map using the $\sqrt{\varepsilon}$ test}
\author{Maria Chikina \footnote{Department of Computational and Systems Biology,
University of Pittsburgh,
3078 Biomedical Science Tower 3,
Pittsburgh, PA 15213,
mchikina@pitt.edu}, 
  Alan Frieze\footnote{
Department of Mathematical Sciences,
Carnegie Mellon University,
Pittsburgh, PA 15213,
alan@random.math.cmu.edu},
  Wesley Pegden\footnote{
Department of Mathematical Sciences,
Carnegie Mellon University,
Pittsburgh, PA 15213,
wes@math.cmu.edu}
}

\usepackage{amsmath,amsthm,amssymb,amscd,latexsym}%,pstricks,fp,calc}

\usepackage{cite,url}
\newcommand{\ep}{\varepsilon}

{
\theoremstyle{definition}

}

{
\theoremstyle{remark}

}

\begin{document}
\maketitle
\begin{abstract}
  In previous work, we developed a rigorous statistical test for outlier status in a reversible Markov Chain, and demonstrated its utilization with an application to detecting gerrymandering in Pennsylvania's Congressional districting.  In this note, we apply our test to the current (Act 43) assembly districting of the state of Wisconsin, and find that the districting is indeed an outlier among the the landscape of valid districtings of Wisconsin.  Outlier status is significant at between $p=.0002$ and $p=.0008$, depending on assumptions.
\end{abstract}
\section{Introduction}
In \cite{outliers}, we developed a new statistical test to rigorously demonstrate outlier status for a state in a reversible Markov Chain; we demonstrated our test with an application to detecting gerrymandering in Pennsylvania's Congressional districts.  In particular, we developed a Markov Chain to sample `valid' districtings of Pennsylvania, and applied our test to show that the present Congressional districting of Pennsylvania is an (extreme) outlier with respect to measures of its Republican bias, for various possible measures of bias and for various possible notions of what constitute valid districtings.

The purpose of the present manuscript is to report the results of an analogous analysis for the assembly districting of the state of Wisconsin.  The structure of the Markov Chain and the statistical test we apply are essentially the same as the one used in \cite{outliers}; thus, in this manuscript, we confine ourselves to describing modifications we made in the chain to reinforce the credibility of the result in the case of Wisconsin, to reporting the results of the test.  (For reproducibility, we describe our data processing steps in Appendix \ref{s.data}.  Our Markov chain code is available at \url{http://math.cmu.edu/~wes/pub.html}.)   We suggest that readers consult our previous paper \cite{outliers}---in particular, the supplementary materials to that paper---for details on the mechanics of our method.

The point of our test is to compare the current districting of Wisconsin with other possible districtings of Wisconsin in a rigorous way.  For example: even though the efficiency gap of Wisconsin is rather large, it might seem conceivable for this efficiency gap to have occurred in a districting ``naturally'', because of the unique political geography of Wisconsin.  Our analysis, which is simple to carry out and grounded in the rigorous statistical framework of \cite{outliers}, shows that the present districting of Wisconsin is an extreme outlier in a quantifiable sense, against the backdrop of other districtings of Wisconsin with similar properties.

\section{Parameters of the Wisconsin chain}

Mechanically, our analysis consists of making many small perturbations of the current Wisconsin districting.  Formally speaking, we are sampling a trajectory from the redistricting Markov chain, which begins with the current districting.  These perturbations are made by choosing a ward on the the boundary of a district, and assigning it instead to a neighboring district, while preserving the properties that
\begin{enumerate}
\item Each district is contiguous,
\item Each district is roughly equal in population (differing from the average district population size by less than 1 average ward population),
\item \label{p.compact} Each district is geometrically reasonable,
\item \label{p.counties} Counties preserved in tact by the Act 43 districting are still preserved,
\item \label{p.mm} Wards in the seven Majority-Minority districts in the Act 43 districting have their district membership preserved.
\end{enumerate}
Property \ref{p.compact} can be formalized in different ways; we have tried 3 natural choices for how to constrain the overall geometry of the districting to be similar to the Act 43 districting (for example, one choice is simply to constrain the total perimeter of all districts in the districting), and find that our analysis is insensitive to the particular choice.  The definitions of these various choices for constraints can be found in the supplementary section of \cite{outliers}.  Districtings which satisfy the above properties are \emph{valid} districtings; our test rigorously compares the Act 43 districting against the landscape of valid districtings of the state of Wisconsin.

\section{Results}

We briefly recall the nature of the $\sqrt \ep$ test.  Repeatedly making perturbations to the Act 43 districting produces a random \emph{trajectory} of districtings, each different from its predecessor by a single ward swap.  We can evaluate the political properties of each districting in the trajectory by carrying out a hypothetical election in each districting using the 2012 Presidential votes in each ward as a proxy for generic Republican/Democrat preference.

The districtings close to the beginning of the trajectory are very similar to the Act 43 districting, and all districtings on the trajectory satisfy properties 1-5, and thus have similar characteristics to the Act 43 districting.  Nevertheless, when we generate a random trajectory of roughly a trillion districtings, the Act 43 districting has a larger efficiency gap than roughly 99.99999\% of districtings on the random trajectory; that is, we find that it's efficiency gap is among the worst $10^{-7}$ fraction seen on the random trajectory.   The finding is relatively insensitive to the choice of how we formalize Property \ref{p.compact}, and also whether we ignore or include Properties \ref{p.counties} and \ref{p.mm} in our analysis, Table \ref{table} summarizes precise results.  

These results show intuitively that the Act 43 districting is favorable to Republicans in carefully crafted way: small changes quickly make the districting fairer.  Common sense dictates that most districtings of Wisconsin shouldn't favor Republicans in such a fragile way, and to confirm this rigorously, the ``$\sqrt{\ep}$ test'' from \cite{outliers} allows us to derive a $p$-value for outlier status of the Act 43 districting on the random trajectories observed in our analysis.   These $p$-values, shown in Table \ref{table}, are rigorously derived upper bounds on the probability that a districting of Wisconsin chosen at random from the set of districtings satisfying properties 1 through 5 (or just 1 through 3; see table headings) would appear as outliers on their trajectories to the degree that the Act 43 districting does.  In particular, our analysis shows that this probability is well under $1/1000$, and our upper bounds are relatively insensitive to how we quantify Property \ref{p.compact}, and whether we include Properties \ref{p.counties} and \ref{p.mm}.

\begin{table}
  \begin{tabular}{c|c|c||c|c}
    Condition for Property \ref{p.compact} & Property \ref{p.counties}? & Property \ref{p.mm}? & $\ep$ & $p$\\
    \hline
    \hline
    Perimeter constraint & yes & yes & $2.7\cdot 10^{-8}$ & $.0002$\\
    L1 constraint & yes & yes & $1.6\cdot 10^{-8}$ & $.0002$\\
    L2 constraint & yes & yes & $1.0\cdot 10^{-8}$ & $.0001$\\
    L1 constraint & no & no & $3.5\cdot 10^{-7}$ & $.0008$\\
  \end{tabular}
  \caption{\label{table}Each line in this table corresponds to a particular random trajectory computed for the Act 43 districting.  The left column indicates the precise geometric condition used; each of these conditions is explained in detail in our full analysis.  The last line is a run without enforcing Properties \ref{p.counties} and \ref{p.mm}; we see good significance here also.  The Fourth column gives the fraction of districtings on the trajectory whose efficiency gaps were as bad as the Act 43 districting.  The final column gives the $p$-values, explained in the text and fully explained in \cite{outliers}. }
\end{table}

\appendix
\section{Data processing notes}
\label{s.data}
Ward data was downloaded from the Wisconsin state Legislative Technology Services Bureau at:\\ \url{ftp://ftp.legis.wisconsin.gov/gis/Website/ElectionData/}\\\url{GIS/Wards_111312_ED_110612.zip}

Our Markov Chain runs on a multigraph abstraction of the ward map.  In order to transform the geographical data into a suitable multigraph we performed the following operations:
\begin{itemize}
  \item Each island precinct was merged with its closest mainland precinct that is in the same assembly district. This operation decreases the number of precincts. 
  \item Multi-polygon precincts were split and their demographic data was split proportionally to the area. This operation increases the number of precincts.
  \item Precincts (or fragments thereof, as generated in Step 2) that are entirely contained within another precinct were removed and their demographic and voting data was assigned to the closest precinct that is also in the same assembly district. This operation decreases the number of precincts.
\end{itemize}
    
The final map has 7292 precincts, starting with 6634 in the original map. This increase comes from splitting of multi-polygon precincts in Step 2; by comparison, changes introduced  step 1 and 3 have negligible effect. All operations preserve district-level voting and demographic data and the efficiency gap in the new map is identical to that of the original map.

\end{document}